\documentclass[lettersize,journal]{IEEEtran}
\IEEEoverridecommandlockouts
\usepackage{cite}
\usepackage{amsmath,amssymb,amsfonts}
\usepackage{algorithm}  
\usepackage{booktabs}
\usepackage{graphicx}
\usepackage{subfigure}
\usepackage[inkscapelatex=false]{svg}
\usepackage{textcomp}
\usepackage{xcolor}
\usepackage{flushend}		
\usepackage{multirow}
\usepackage{booktabs}
\usepackage{svg}
\svgsetup{
    inkscapepath=i/svg-inkscape/
}
\svgpath{{svg/}}

\usepackage{algpseudocode}  

\usepackage{comment}

\usepackage{tikz}
\usepackage{pifont}

\usepackage{mathtools, nccmath}
\usepackage{xpatch}
\xpatchcmd{\NCC@ignorepar}{%
\abovedisplayskip\abovedisplayshortskip}
{%
\abovedisplayskip\abovedisplayshortskip%
\belowdisplayskip\belowdisplayshortskip}
{}{}

\usepackage{cases}

\usepackage{bbm}

\usepackage{lipsum}

\usepackage{amsmath}

\usepackage{pict2e}
\newsavebox{\ORCIDlogo}
\savebox{\ORCIDlogo}{%
\setlength{\unitlength}{\dimexpr 1em/256\relax}%
\begin{picture}(256,256)%
  \color[HTML]{A6CE39}\put(128,128){\circle*{256}}%
  \color{white}%
  \put(78.6,199.2){\circle*{20}}%
  \moveto(70.9,176,9)\lineto(86.3,176,9)\lineto(86.3,69.8)\lineto(70.9,69.8)%
  \closepath\fillpath%
  \moveto(108.9,176.9)\lineto(150.5,176.9)%
  \curveto(190.1,176.9)(207.5,148.6)(207.5 ,123.3)%
  \curveto(207.5,95,8)(186,69.7)(150.7,69.7)%
  \lineto(108.9,69.7)%
  \closepath\fillpath%
  \color[HTML]{A6CE39}%
  \moveto(124.3,83.6)\lineto(148.8,83.6)%
  \curveto(183.7,83.6)(191.7,110.1)(191.7,123.3)%
  \curveto(191.7,144.8)(178,163)(148,163)%
  \lineto(124.3,163)%
  \closepath\fillpath%
\end{picture}%
}

\newcommand\orcidicon[1]{\href{https://orcid.org/#1}{\usebox{\ORCIDlogo}}}

\usepackage[colorlinks=false, urlcolor=black, pdfborder={0 0 0}]{hyperref}

\def\BibTeX{{\rm B\kern-.05em{\sc i\kern-.025em b}\kern-.08em
    T\kern-.1667em\lower.7ex\hbox{E}\kern-.125emX}}

\usepackage{float}
\usepackage{threeparttable}
\usepackage{acronym}
\newacro{KV}{key-value}
\newacro{LLM}{large language model}
\newacro{CIM}{compute-in-memory}
\newacro{LoRA}{low-rank adaptation}
\newacro{PE}{processing element}
\newacro{IPCN}{Inter-PE computational network}
\newacro{SRPG}{SRAM reprogramming and power gating}
\newacro{CT}{compute tile}
\newacro{DMAC}{multiply-accumulate operations on dynamic data}
\newacro{SMAC}{multiply-accumulate operations on static weights}
\newacro{RRAM-ACIM}{resistive RAM analog compute-in-memory}
\newacro{SRAM-DCIM}{static RAM digital compute-in-memory}
\newacro{NMC}{network main controller}
\newacro{NPM}{Network Program Memory}
\newacro{OS FeFET}{oxide semiconductor ferroelectric field-effect transistor}
\newacro{NVM}{non-volatile memory}
\newacro{BEOL}{back‑end‑of‑line}
\newacro{FEOL}{front‑end‑of‑line}
\newacro{M3D}{monolithic 3D}
\newacro{ADC}{analog‑to‑digital converter}
\newacro{DRAM}{dynamic RAM}
\newacro{MVM}{matrix-vector multiplication}
\newacro{SAR}{successive approximation register}
\newacro{ISA}{instruction set architecture}
\newacro{eDRAM}{embedded DRAM}
\newacro{FIFO}{First-In First-Out buffer}
\newacro{ISA}{instruction set architecture}
\newacro{API}{application programming interface}
\newacro{2T0C}{two-transistor zero-capacitor}
\newacro{2D mesh}{two-dimensional mesh}
\newacro{UCIe}{Universal Chiplet Interconnect Express}
\newacro{TTFT}{Time-to-first-token}
\newacro{ITL}{Inter-token latency}

\usepackage{cleveref}
\crefname{figure}{fig.}{figs.}

\begin{document}

\title{CHIPSMORE: Compute-in-Interconnect and -Memory \underline{Chip}lets for Multi-\underline{Mo}de Multi-\underline{Re}quest LLM Inference Acceleration\\
}

\author{\IEEEauthorblockN{Yue Jiet Chong\orcidicon{0009-0001-7679-4252}, Yimin Wang\orcidicon{0009-0008-7292-0670}, Zhen Wu\orcidicon{0009-0005-4644-0903}, Zixuan Wang\orcidicon{0009-0006-9486-3211}, Wei Zhang\orcidicon{0009-0001-5331-3137}, Xuanyao Fong\orcidicon{0000-0001-5939-7389},~\IEEEmembership{Member,~IEEE}}

\thanks{This work is funded in part by the National University of Singapore through the Microelectronics Seed Grant (FY2024); and in part by A*STAR under the RIE2030 Energy-aware Accelerated Computing program (Award H25-MSR3439). \textit{(corresponding author: Xuanyao Fong)}
}

\thanks{The authors are with the Department of Electrical and Computer Engineering, National University of Singapore, Singapore. E-mail: \{jason.yj.chong, kelvin.xy.fong\}@nus.edu.sg
}

}


\maketitle

\begin{abstract}
\Ac{LLM} inference exhibits substantial variability across adaptation modes, context lengths, and request concurrency, creating challenges for maintaining high utilization, memory efficiency, and scalable performance on \ac{CIM} accelerators.
This paper presents \textit{CHIPSMORE}, a multi-mode and multi-request LLM inference accelerator that integrates compute-in-interconnect and \ac{CIM} to support both base-mode and \ac{LoRA} inference under diverse workloads.
\textit{CHIPSMORE} employs heterogeneous processing elements consisting of \ac{RRAM-ACIM} and \ac{SRAM-DCIM} interconnected through a programmable \ac{IPCN}.
A composable hierarchical \ac{KV} memory scheme dynamically allocates router scratchpad, \ac{SRAM-DCIM}, and \ac{eDRAM} resources according to workload requirements, enabling scalable support for long-context and batched inference.
Furthermore, a non-replicated multi-request execution pipeline exploits request-level parallelism without duplicating pretrained weights, while a state-aware resource reconfiguration mechanism selectively retains runtime states and power-gates inactive resources to improve energy efficiency. 
Evaluation using cycle-accurate hardware-software co-simulation demonstrates that \textit{CHIPSMORE} effectively sustains high throughput across varying model sizes, context lengths, and batch sizes while maintaining favorable power scaling.
Compared with Nvidia H100, \textit{CHIPSMORE} achieves up to $2.38\times$ higher throughput and $27\times$ higher energy efficiency on Mistral-7B inference while eliminating weight replication for multi-request serving.

\end{abstract}


\section{Introduction}

\Acp{LLM} have demonstrated strong capabilities in natural-language understanding, content generation, code synthesis, and domain-specific reasoning, leading to their increasing deployment in both cloud and edge computing systems~\cite{llm_in_industry}.
However, the computational and memory demands of \ac{LLM} inference continue to grow with model size, context length, and service concurrency~\cite{llm_trend}.
\ac{LLM} execution consists of two distinct phases.
The \textit{prefill} phase processes the input prompt with matrix-matrix computation and parallelism, whereas the \textit{decode} phase generates tokens autoregressively and repeatedly accesses the \ac{LLM} weights and \ac{KV} cache~\cite{vllm}.
Consequently, an accelerator optimized for prefill computation may remain underutilized during decode, while a design provisioned for long-context decode may incur excessive area and static power when serving short-context workloads~\cite{pim_gpt, attacc, neupims}.

The hardware requirements of \ac{LLM} inference depend on request batch size and adaptation mode.
Meanwhile, the \ac{KV}-cache capacity grows linearly with batch size and retained context length~\cite{kv_cache_scaling}.
Increasing the batch size can improve aggregate throughput by exposing request-level parallelism, but it also multiplies the volatile \ac{KV} state that must be stored and preserved~\cite{llm_multi_request}.
Base-mode inference primarily accesses static pretrained weights, which were obtained by training the \ac{LLM} on specific tasks.
To improve the performance of \ac{LLM} on new tasks without retraining, \ac{LoRA}~\cite{lora_origin} trains a small set of task-specific matrices (called \textit{adapters}) on those tasks.
The adapters are used in place of targeted pretrained weights when the \ac{LLM} works on the new task.
Thus, the broad operating space of \ac{LLM} poses a challenge to optimizing the hardware \ac{LLM} inference accelerator design to satisfy the broad range of requirements.

\begin{figure}[t]
    \centering
    \includegraphics[width=0.75\linewidth]{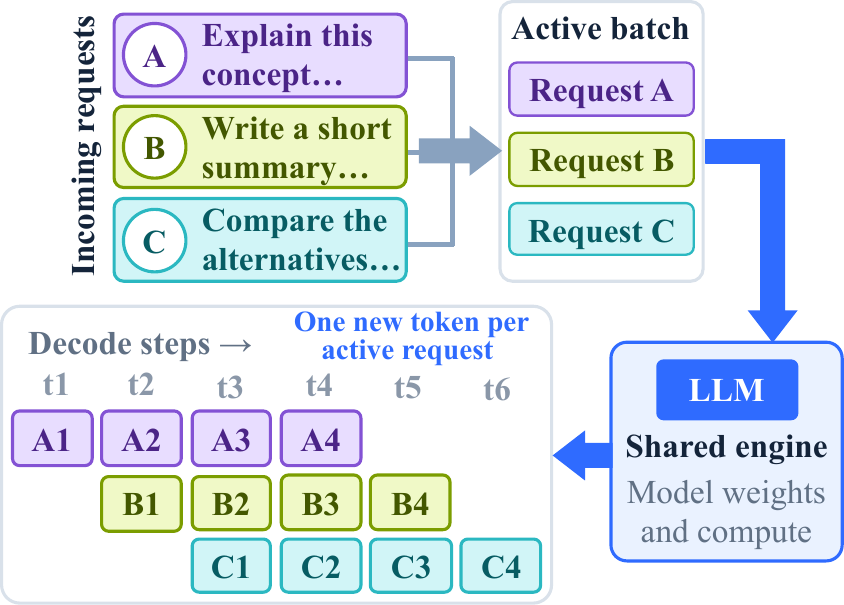}
    \caption{Illustration of multi-request LLM inference.}
    \label{fig:llm_batching_high_level}
\end{figure}

\Acf{CIM} architectures offer a promising approach to reducing the energy and latency associated with repeated weight movement~\cite{nonVN_survey, jetcas_jade, cim_survey}.
Nonvolatile \ac{RRAM-ACIM} is particularly suitable for storing and processing static pretrained weights due to its density, nonvolatility, and high matrix-vector parallelism~\cite{rram_survey}.
However, the limited write endurance and high write energy of \acp{RRAM-ACIM} make them unsuitable for processing dynamic intermediate data that is generated during \ac{LLM} inference \cite{cim_survey}.
In contrast, \ac{SRAM-DCIM} provides rapid reprogrammability and deterministic digital computation~\cite{sram_survey}, making it suitable for \ac{LoRA} matrices and temporary storage.
Nevertheless, existing \ac{CIM} accelerators commonly assume a fixed context, a single memory role, or single-batch execution~\cite{pim_gpt, attacc, neupims, cent, h2_llm, chime, iscas_primal}.
Designs that improve batch throughput through weight replication increase RRAM capacity, chiplet count, area, and leakage approximately in proportion to the replication factor~\cite{weight_replicate_1, weight_replicate_2}.
Moreover, designs relying exclusively on local SRAM for \ac{KV} storage face a different scalability limitation because the required capacity grows rapidly with context length and batch size~\cite{sram_kv_limit_1, transpim}.

This article presents \textit{CHIPSMORE}, a multi-mode and multi-request chiplet-based \ac{CIM} architecture supporting both base and \ac{LoRA} \ac{LLM} inference on various context lengths and batch sizes.
The main contributions are as follows.

\begin{itemize}
    \item A workload-oriented orchestration framework that unifies adaptation-aware execution, \ac{KV}-cache scaling, and multi-request serving for \ac{CIM}-based \ac{LLM} inference.
    
    \item A persistence-aware compute partitioning architecture that maps static-weight operations to heterogeneous \ac{CIM} macros and runtime-generated operations to a programmable \ac{IPCN}, enabling communication and computation co-execution on dynamic inference data.
    
    \item A composable hierarchical \ac{KV} memory scheme that dynamically combines and allocates router scratchpad, \ac{SRAM-DCIM}, and integrated \ac{eDRAM} according to workload mode and capacity requirements.

    \item A non-replicated multi-request pipeline that executes multiple inference requests across unique weight-bearing chiplet clusters without replicating RRAM weights or increasing RRAM area with batch size.

    \item A state-aware chiplet clustering and power gating scheme that keeps weight-bearing clusters active, retains clusters containing volatile \ac{KV} or \ac{LoRA} state, and powers off stateless resources where possible to enable sub-linear power scaling with respect to batch and model sizes.

\end{itemize}

\section{Background and Motivation}

\begin{table*}[t] 
    \caption{Comparison of workload adaptability of existing LLM accelerators}  
    \centering \footnotesize 
    \setlength{\tabcolsep}{1pt}
    \begin{tabular}{lcccccccc} 
        \toprule 
        \textbf{Feature} & \textbf{PIM-GPT \cite{pim_gpt}} & \textbf{AttAcc \cite{attacc}} & \textbf{NeuPIMs \cite{neupims}} & \textbf{CENT \cite{cent}} & \textbf{H2-LLM \cite{h2_llm}} & \textbf{CHIME \cite{chime}} &\textbf{PRIMAL \cite{iscas_primal}} & \textbf{This Work} \\
        \midrule 
        Base-Mode Inference & \checkmark & \checkmark & \checkmark & \checkmark & \checkmark & \checkmark & \checkmark & \checkmark \\ 
        
        LoRA Adaptation Support & $\times$ & $\times$ & $\times$ & $\times$ & $\times$ & $\times$ & \checkmark & \checkmark \\ 
        
        Dynamic KV-Capacity Scaling & $\times$ & $\times$ & \checkmark & $\times$ & $\times$ & \checkmark & $\times$ & \checkmark \\ 
        
        Hierarchical KV Memory Management & $\times$ & $\times$ & $\times$ & $\times$ & $\times$ & $\times$ & $\times$ & \checkmark \\ 
        
        Multi-Request Inference & $\times$ & \checkmark & \checkmark & \checkmark & \checkmark & \checkmark & $\times$ & \checkmark \\ 
        
        Weight Replication-Free Throughput Scaling & $\times$ & $\times$ & $\times$ & \checkmark & $\times$ & $\times$ & $\times$ & \checkmark \\ 
        
        State-Aware Power Management & $\times$ & $\times$ & $\times$ & $\times$ & $\times$ & $\times$ & $\times$ & \checkmark \\ 
        
        Joint Adaptation / Context / Batch Optimization & $\times$ & $\times$ & $\times$ & $\times$ & $\times$ & $\times$ & $\times$ & \checkmark \\ 
        \bottomrule 
    \end{tabular} 
    
    \label{tab:comparison} 
\end{table*}

\subsection{Low-Rank Adaptation (LoRA)}
\ac{LoRA} \cite{lora_origin} augments a pretrained model with a small set of task-specific low-rank matrices while keeping the original model parameters unchanged.
During inference, the pretrained weights are shared across different downstream tasks, whereas only the adaptation matrices need to be updated when switching between tasks.
This separation creates distinct storage requirements for the two classes of parameters.

For \ac{CIM}-based accelerators, pretrained weights are well suited to non-volatile memory due to their persistent nature, while \ac{LoRA} parameters benefit from memory resources that support rapid reconfiguration \cite{ahwa_lora, iscas_primal}.
Consequently, supporting both base-mode and \ac{LoRA} inference requires a heterogeneous memory organization that can efficiently accommodate static and reprogrammable parameters simultaneously. 
This observation motivates the combined use of \ac{RRAM-ACIM} and \ac{SRAM-DCIM} resources in \textit{CHIPSMORE}.

\subsection{Multi-Request Inference}
Multi-request inference improves accelerator utilization by serving multiple independent requests concurrently on shared hardware resources.
While the prefill phase typically exposes sufficient intra-request parallelism through prompt processing, the decode phase generates only one token per request at each autoregressive step, resulting in limited parallelism for a single request.
Consequently, batching multiple requests becomes an effective approach for increasing hardware utilization and throughput during decoding \cite{vllm}.
In addition, all requests served by the same model access an identical set of pretrained parameters, creating an opportunity to reuse mapped model weights across requests.

However, supporting multi-request execution on \ac{CIM} accelerators introduces two key challenges.
First, each request maintains an independent \ac{KV} cache whose capacity grows with retained context length, causing memory demand to increase with request concurrency.
Second, existing throughput-scaling approaches often rely on replicating model weights across multiple processing instances, which proportionally increases storage capacity, area, and power consumption despite the weights being identical across requests \cite{weight_replicate_1}.
An efficient multi-request architecture should therefore improve throughput while preserving weight sharing and providing scalable support for request-specific runtime states.

\subsection{KV-Cache Scaling}

\begin{figure}[t]
    \centering
    \includegraphics[width=0.7\linewidth]{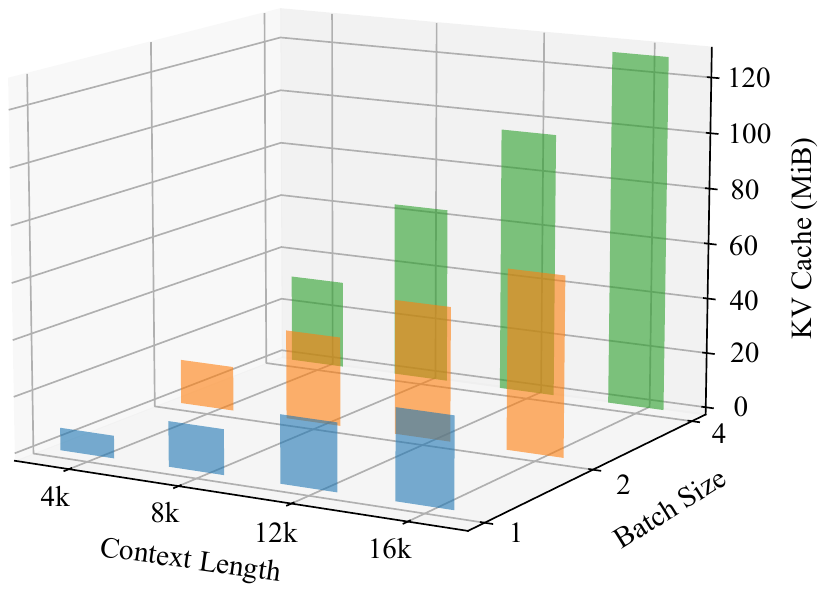}
    \caption{KV cache capacity versus context length and batch size (Llama 3.2-1B).}
    \label{fig:kv_capacity_vs_batch_context}
\end{figure}

During \ac{LLM} inference, each active request maintains a \ac{KV} cache that stores attention states generated by previously processed tokens.
Unlike model parameters, which remain fixed during inference, \ac{KV}-cache capacity grows with both retained context length and request concurrency. 
Consequently, \ac{KV} storage becomes a significant consumer of on-chip memory resources, particularly for long-context and multi-request workloads.

For a homogeneous batch of $B$ requests with a retained context of $T$ tokens, the required \ac{KV}-cache capacity is \\

\begin{equation}
C_{KV} = BTL(2N_{KV}d_{head})\frac{b_{KV}}{8} \text{ $bytes$}
\label{eq:einstein}
\end{equation}
\newline
where \(L\) is the number of attention layers, \(N_{\mathrm{KV}}\) is the number of K/V heads, \(d_{\mathrm{head}}\) is the head dimension, and \(b_{\mathrm{KV}}\) is the number of bits used for each K/V element.
The factor of two accounts for both key and value vectors. 
As context length and request concurrency increase, the aggregate \ac{KV}-cache requirement can exceed the capacity of local memory resources, making scalable memory allocation a primary design challenge for \ac{CIM}-based \ac{LLM} accelerators \cite{transpim}.
\Cref{fig:kv_capacity_vs_batch_context} shows the KV capacity across different context lengths and batch size.
The rapid growth in memory demand highlights the need for a flexible memory hierarchy capable of accommodating workload-dependent \ac{KV}-storage requirements while preserving data locality during inference.

\subsection{Research Gap and Design Opportunity}
The preceding discussion highlights three key challenges in \ac{LLM} inference acceleration: adaptation-mode variability introduced by \ac{LoRA}, \ac{KV}-cache capacity growth with context length, and throughput scaling through concurrent request execution.
Although existing \ac{CIM}-based accelerators (Table I) have demonstrated high efficiency for specific workloads, most target only one or two of these dimensions.
As a result, adaptation support, \ac{KV}-memory management, and multi-request execution are typically addressed independently, leading to suboptimal resource utilization when workload characteristics change.

Furthermore, the underlying compute, memory, and communication resources are strongly coupled across these workload dimensions.
For example, memory resources allocated for adaptation support may remain underutilized during base-mode inference, while throughput-oriented approaches often increase hardware cost by replicating pretrained weights despite all requests sharing the same model parameters.
Meanwhile, \ac{KV}-cache demand varies significantly with both context length and request concurrency, requiring memory resources to be allocated dynamically according to runtime requirements.

These observations motivate a unified \ac{CIM} architecture that jointly orchestrates computation, communication, memory allocation, and runtime-state management across diverse operating conditions.
The proposed \textit{CHIPSMORE} architecture is presented in Section III.

\section{\textit{CHIPSMORE} Hardware Architecture}

\begin{figure*}[t]
    \centering
    \includegraphics[width=1\linewidth]{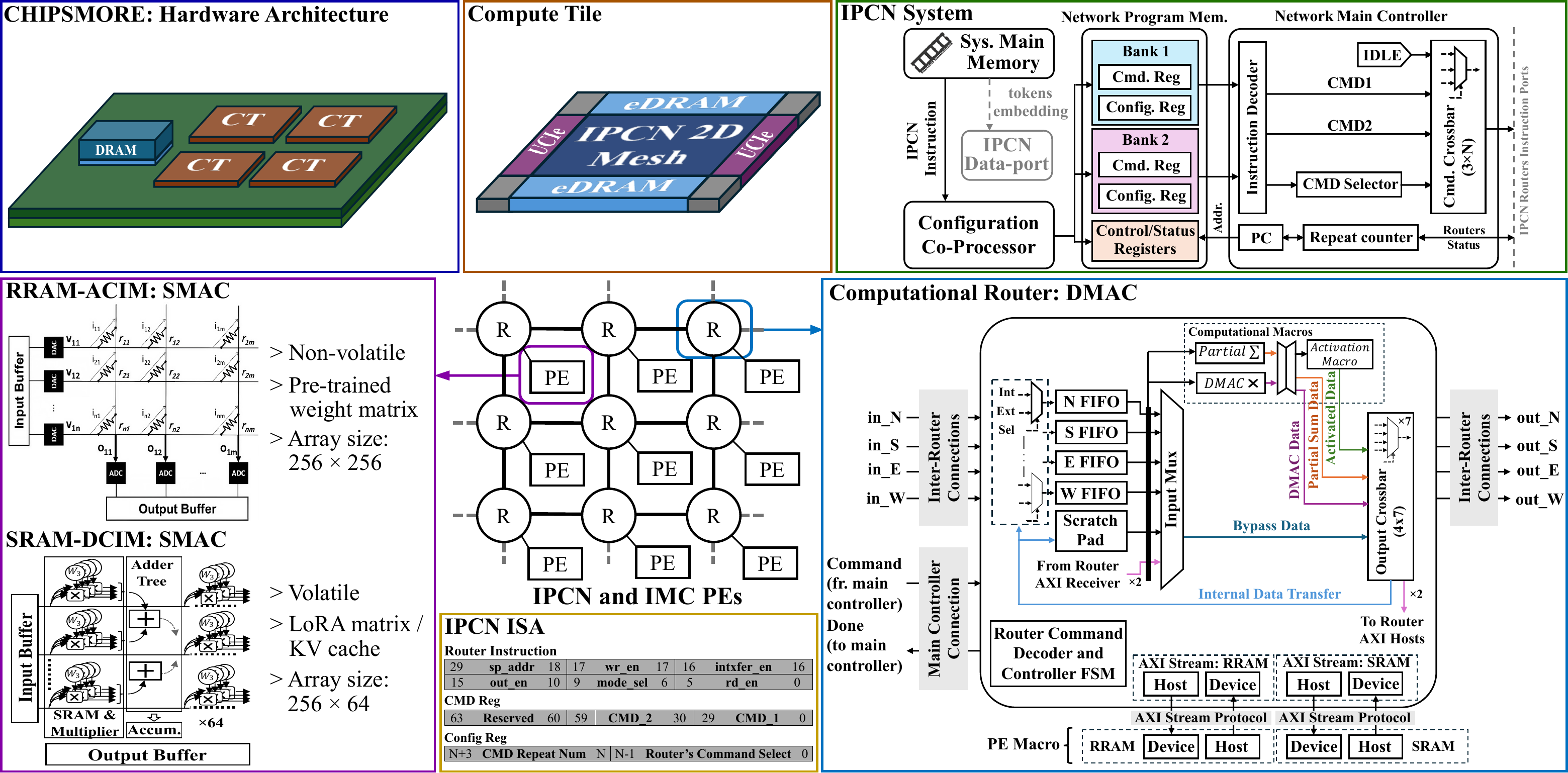}
    \caption{CHIPSMORE hardware architecture: heterogeneous \ac{PE}, 2D-Mesh \ac{IPCN}, computational router.}
    \label{fig:hardware_arch}
\end{figure*}

To efficiently support diverse \ac{LLM} inference scenarios, including base-mode and LoRA execution as well as varying context lengths and request-level parallelism, \textit{CHIPSMORE} adopts a hierarchical chiplet-based architecture.
The system is composed of multiple \acp{CT}, each serving as an autonomous processing cluster that integrates computation, communication, and storage resources.
Within a CT, heterogeneous \acp{PE} are interconnected through a \ac{2D mesh} \ac{IPCN}, as illustrated in \Cref{fig:hardware_arch}.
The \ac{IPCN} not only provides data communication among distributed \ac{CIM} resources but also performs in-network computations on dynamic runtime data, such as the attention score computation $\mathbf{Q} \cdot \mathbf{K^T}$, referred to as \ac{DMAC} operations.
In parallel, the \acp{PE} execute \ac{SMAC} operations on model parameters stored within the \ac{CIM} arrays, corresponding to the matrix-vector multiplications that dominate Transformer~\cite{transformer_arch} inference workloads.
This partitioning of computation between the \ac{IPCN} and \acp{PE} enables efficient execution of both data-dependent and weight-dependent operations while minimizing data movement across the system.

Unlike conventional \ac{CIM} accelerators, which typically execute both weight-dependent and runtime-generated operations using the same computational substrate, \textit{CHIPSMORE} partitions computation according to operand persistence.
Operations associated with static pretrained parameters are executed within the \ac{CIM} macros through \ac{SMAC} execution, whereas operations involving runtime-generated activations and intermediate tensors are offloaded to the \ac{IPCN} through \ac{DMAC} execution.
This persistence-aware compute partitioning enables each hardware resource to be specialized according to the characteristics of the operands it processes, reducing activation movement while preserving the weight-stationary benefits of \ac{CIM} execution.

\subsection{Processing Element (PE)}
The \ac{PE} is the fundamental computational unit of \textit{CHIPSMORE}.
Each \ac{PE} integrates a non-volatile \ac{RRAM-ACIM} macro and a volatile \ac{SRAM-DCIM} macro, enabling efficient support for both frozen pretrained weights and reconfigurable \ac{LoRA} parameters.
By combining dense non-volatile analog computing with programmable digital \ac{CIM}, the \ac{PE} provides a heterogeneous computing substrate that accommodates the distinct storage, precision, and update requirements of modern LLM workloads.
Within the \ac{PE}, both \ac{CIM} macros perform \ac{SMAC} operations on locally stored weights, while the choice of memory technology is determined by the persistence and reconfiguration characteristics of the target parameters.

\subsubsection{RRAM-ACIM}
The \ac{RRAM-ACIM} macro~\cite{rram_acim} serves as the primary storage and computation engine for pretrained model parameters.
Owing to its high storage density and non-volatility, RRAM is well suited for accommodating the large weight matrices of foundation models that remain unchanged during inference.
Once programmed, the weights can be retained without refresh or reconfiguration overhead, allowing the base model to remain persistently mapped onto the accelerator.
During execution, \acp{MVM} are performed directly within the analog crossbar arrays, exploiting the intrinsic parallelism of in-memory computing to achieve high throughput and energy efficiency.
As a result, \ac{RRAM-ACIM} provides an effective platform for executing the dominant weight-stationary operations of Transformer inference while minimizing data movement between memory and compute resources.

\subsubsection{SRAM-DCIM}
The \ac{SRAM-DCIM} macro~\cite{sram_dcim} complements the \ac{RRAM-ACIM} by supporting reconfigurable model parameters.
Although SRAM provides lower storage density than RRAM, its fast write capability enables efficient deployment of task-specific adaptation parameters. 
Consequently, \ac{SRAM-DCIM} is used to store \ac{LoRA} matrices and execute the corresponding digital MAC operations without modifying the pretrained
weights permanently mapped onto the \ac{RRAM-ACIM} arrays.

In addition to its role in adaptation support, the \ac{SRAM-DCIM} resource remains accessible to the \ac{IPCN} and can function as a general-purpose on-chip storage resource when adaptation parameters are absent.
This flexibility enables \ac{SRAM-DCIM} to serve both computational and storage roles within the \ac{PE} while maintaining a unified hardware structure across different operating modes.

\subsection{Inter-PE Computational Network (IPCN)}
The \ac{IPCN} serves as the communication and orchestration backbone of \textit{CHIPSMORE}, interconnecting the distributed \ac{CIM} resources within a \ac{CT}.
Beyond conventional data transport, the \ac{IPCN} incorporates in-network computing capabilities that enable selected operations to be executed directly during data movement, thereby reducing communication overhead and improving data locality.
Computations on runtime-generated activations and intermediate tensors, referred to as \ac{DMAC} operations, are executed within the \ac{IPCN} routers.
Executing \ac{DMAC} operations in-network eliminates intermediate activation transfers to discrete processing engines and preserves locality between communication and computation.
As illustrated in \Cref{fig:hardware_arch}, the IPCN comprises three principal components: the \ac{NPM}, the \ac{NMC}, and a \ac{2D mesh} of computational routers with \acp{PE}. 

\subsubsection{Network Program Memory (NPM)}
The \ac{NPM} serves as the instruction storage unit of the \ac{IPCN} and enables software-defined control of data movement and in-network computation, allowing the network behavior to be reconfigured according to the communication and computational requirements of different AI workloads.

The \ac{NPM} is organized into three banks: Bank 1 (B1), Bank 2 (B2), and a Control and Status Register (CSR) bank.
Both B1 and B2 comprise two sub-banks, namely the Command Register (CMR) and the Configuration Register (CFR).
The CMR stores the router commands associated with a program step, while the CFR specifies the command-selection mask and execution count for each router within the \ac{IPCN}. 
For every instruction row, two candidate commands (CMD1 and CMD2) are stored in the CMR, whereas the corresponding CFR entry determines whether a router executes CMD1, CMD2, or remains idle.
The CFR additionally records the number of repetitions for the selected operation, allowing a single instruction entry to represent repetitive communication/computation patterns and thereby reducing program storage overhead.

\subsubsection{Configuration Co-processor (CC)}
The CC is responsible for managing \ac{NPM} updates and enabling runtime reconfiguration of the \ac{IPCN}.
It loads program instructions generated by the software toolchain from the system memory into the \ac{NPM} without interrupting ongoing execution.
By decoupling program loading from instruction execution, CC allows the \ac{IPCN} to adapt its dataflow and computation schedule to different workloads while reducing configuration overhead.

To sustain continuous operation, the \ac{NPM} employs a double-buffered organization using B1 and B2.
At any time, one bank serves as the active instruction source for the \ac{NMC}, while the other bank is updated by the CC.
Specifically, when the \ac{NMC} fetches instructions from one bank, the CC concurrently populates the other bank with the next instruction sequence.
This bank-interleaved execution scheme overlaps instruction loading with network execution, thereby minimizing reconfiguration-induced stalls and reducing \ac{IPCN} idle cycles.

\subsubsection{Network Main Controller (NMC)}
The \ac{NMC} serves as the centralized control unit of the \ac{IPCN}, coordinating the operation of distributed routers and \acp{PE} within a \ac{CT}.
During execution, the \ac{NMC} sequentially fetches instructions from the active \ac{NPM} bank, decodes the instruction fields, and generates the control signals required to establish the desired communication and computation patterns across the network.
To support efficient execution of repetitive communication and computation patterns, the \ac{NMC} also manages instruction iteration and execution sequencing.

By centralizing instruction sequencing and command distribution, the \ac{NMC} enables synchronized execution of routing, local-memory access, and DMAC operations throughout the \ac{IPCN}.
This control mechanism allows the programmable network to realize workload-specific dataflow patterns while maintaining a uniform hardware structure.

\subsubsection{Unit Router}
The unit router is the fundamental building block of the \ac{IPCN} and serves as both a communication node and an in-network computing engine.
Organized in a \ac{2D mesh} topology, the routers collectively establish the data transport fabric of the \ac{CT} while enabling computation to be performed directly on data residing within the network.
Each router is coupled with a pair of heterogeneous \acp{PE}, forming a router-PE pair that jointly supports both DMAC and SMAC execution.

The unit router consists of the following sub-modules: 
(i) \textit{Data I/O ports}: 4 planar ports for inter-router connections and a pair of AXI-Stream adapters for router-\acp{PE} connection.
Each port is integrated with \ac{FIFO} for temporary data storage, enabling concurrent data transfers across the network.
(ii) \textit{Controller}: Based on the decoded instruction, the controller coordinates packet routing, local memory accesses, internal data transfers, and computational operations within the router.
(iii) \textit{Computational Macros}: These macros support operations frequently encountered in AI workloads, including partial sums, activations, and \acp{DMAC}.

\subsubsection{IPCN Instruction Set Architecture (ISA)}
To enable software-defined control of communication and in-network computation, the \ac{IPCN} employs a dedicated \ac{ISA}.
By exposing network functionality through programmable instructions rather than fixed control logic, the \ac{IPCN} can adapt its execution behavior to different AI workloads while maintaining a uniform hardware structure.

Each \ac{IPCN} instruction is encoded as a 30-bit control word, as shown in \Cref{fig:hardware_arch}, and contains the information required to coordinate routing, computation, and local memory operations across the network.
The \ac{ISA} supports both communication primitives, including unicast and broadcast transmission, and computation primitives executed within the routers. 
Through this programmable interface, the \ac{IPCN} can realize a wide range of workload-specific dataflow patterns without hardware modification.
To facilitate software development, an accompanying programming toolchain comprising an \ac{API} and a compiler is provided.
The toolchain translates high-level descriptions of communication and computation schedules into executable \ac{IPCN} instruction sequences.

\subsection{Embedded DRAM (eDRAM)}
Each CT integrates a multi-bank \ac{eDRAM} block that provides a high-density on-chip storage resource.
Compared with router scratchpad memory and \ac{SRAM-DCIM}, \ac{eDRAM} offers substantially higher storage capacity within a given silicon area, making it suitable for capacity-oriented storage requirements that exceed the resources available in local SRAM structures.

The \ac{eDRAM} is implemented using a \ac{2T0C} cell structure \cite{edram}.
Unlike conventional 1T1C \ac{eDRAM}, the capacitorless design simplifies integration with advanced logic processes and improves scalability.
Furthermore, the \ac{2T0C} cell supports non-destructive read operations, eliminating the need for data restoration after each read access and reducing memory access overhead.
The longer retention time of the cell also lowers refresh frequency, improving the energy efficiency of storing large amounts of runtime inference state.
Consequently, \ac{eDRAM} serves as a high-capacity memory resource that complements the lower-latency router scratchpad and \ac{SRAM-DCIM} resources within each \ac{CT}.

\subsection{Inter-CT Communication}
To scale beyond the resource limits of a single \ac{CT}, \textit{CHIPSMORE} adopts a chiplet-based architecture in which multiple \acp{CT} are interconnected through a die-to-die communication fabric.
While the \ac{IPCN} provides low-latency communication within a \ac{CT}, inter-tile communication is required to support larger models, expanded memory capacity, and distributed execution across multiple chiplets.

Inter-\ac{CT} communication is implemented using the \ac{UCIe} standard \cite{ucie}.
Each \ac{CT} integrates a \ac{UCIe} endpoint that provides a high-bandwidth and low-latency communication interface to neighboring chiplets.
The standardized die-to-die interface enables efficient exchange of activations, intermediate tensors, and memory data between \acp{CT} while maintaining compatibility with heterogeneous chiplet integration technologies.

To minimize communication overhead, computation and storage resources are preferentially localized within individual compute tiles whenever possible. 
Inter-tile transfers are therefore limited to workload-dependent data exchanges that cannot be satisfied by the local \ac{IPCN} and memory hierarchy.
This design preserves the high-efficiency data locality enabled by the \ac{CIM}-based architecture while allowing the system to scale to larger models, longer context lengths, and higher request concurrency through the addition of compute tiles.
From the perspective of the software runtime, the collection of \ac{CT} forms a unified accelerator fabric, while \ac{UCIe} transparently transports data between physically distributed resources.
\section{Workload-Oriented Resource Orchestration}

\begin{table}[t]
    \caption{CHIPSMORE Modes of Operation}
    \centering
    \setlength{\tabcolsep}{4pt}
    \begin{tabular}{c|c|c|c|c}
         \hline
         \multirow{2}{*}{Mode} & Batch & Context Tokens & Weight & KV-allocation Priority \\
         & Size & (Input/Output) & Placement & (Highest $\rightarrow$ Lowest)\\
         \hline
         \multirow{3}{*}{Base} & \multirow{3}{*}{1, 2, 4} &
         S: 2048/2048 & \multirow{3}{*}{\shortstack{Base: RRAM}} & Router Scratchpad $\rightarrow$ \\
         & & L: 4096/4096 & & SRAM-DCIM $\rightarrow$ \\
         & & XL: 8192/8192 & & eDRAM \\
         \hline
         \multirow{3}{*}{LoRA} & \multirow{3}{*}{1, 2, 4} &
         S: 2048/2048 & \multirow{3}{*}{\shortstack{Base: RRAM \\ LoRA: SRAM}} & \multirow{3}{*}{\shortstack{Router Scratchpad $\rightarrow$\\ eDRAM}} \\
         & & L: 4096/4096 & & \\
         & & XL: 8192/8192 & & \\
         \hline
    \end{tabular}
    \label{tab:op_mode}
\end{table}

Efficient support for heterogeneous \ac{LLM} workloads requires additional mechanisms beyond static weight placement.
In practice, inference workloads vary across adaptation modes (base model and \ac{LoRA}), context lengths, and request batch sizes, resulting in substantially different requirements for memory capacity, data locality, and resource utilization.
To address this challenge, \textit{CHIPSMORE} introduces a workload-oriented resource orchestration framework comprising four key components: Unified Spatial Mapping, Hierarchical KV Memory Allocation, Non-replicated Multi-request Pipeline, and State-aware Resource Reconfiguration.
Together, these techniques enable \textit{CHIPSMORE} to efficiently support a broad operating range spanning base-mode inference, \ac{LoRA} adaptation, long-context execution, and multi-request serving while maintaining high hardware utilization and energy efficiency.

\subsection{Unified Spatial Mapping}

\textit{CHIPSMORE} adopts the spatial mapping methodology developed in LEAP~\cite{iccad_leap} and PRIMAL~\cite{iscas_primal} to preserve locality between computation and data storage while enabling a uniform execution model across base and \ac{LoRA} inference modes.
The pretrained projection matrices $\mathbf{W_Q}$,$\mathbf{W_K}$,$\mathbf{W_V}$, and $\mathbf{W_O}$ 
are partitioned according to the dimensions of the \ac{CIM} crossbar arrays and spatially mapped onto \ac{RRAM-ACIM} \acp{PE}.
In parallel, the corresponding \ac{LoRA} matrices are partitioned using the same geometric decomposition and placed in the \ac{SRAM-DCIM}.
By maintaining identical partition boundaries and spatial correspondence between pretrained and adaptation weights, the \ac{LoRA} contribution can be evaluated locally and accumulated with the base model projection output.

To simplify routing and maintain balanced \ac{IPCN} traffic, each partitioned weight matrix is constrained to occupy a contiguous column-wise rectangular region.
The resulting layout aligns communication paths with the regular structure of the 2D-mesh \ac{IPCN}, allowing reductions and broadcasts to be performed along deterministic horizontal and vertical routes with minimal congestion.
Since \ac{LoRA} matrices share the same dimensional structure as their corresponding pretrained matrices, no additional placement constraints are required when switching between inference modes.

Intermediate tensors ($\mathbf{Q}$, $\mathbf{K}$, $\mathbf{V}$, and $\mathbf{O}$) are stored in the distributed memory resources associated with the router-PE pairs containing the corresponding weight partitions.
This locality-aware placement minimizes communication overhead by allowing partial-result aggregation and subsequent tensor accesses to be performed within the same spatial region.
The spatial organization remains invariant across all operating modes of \textit{CHIPSMORE}.
Consequently, workload adaptation does not require remapping of the \acp{PE}, and only the allocation of runtime state within the memory hierarchy is modified according to workload requirements.
This separation between static weight mapping and dynamic resource allocation forms the foundation for the hierarchical memory management and multi-request execution mechanisms described in the following subsections.

\begin{figure}[t]
    \centering
    \includegraphics[width=\linewidth]{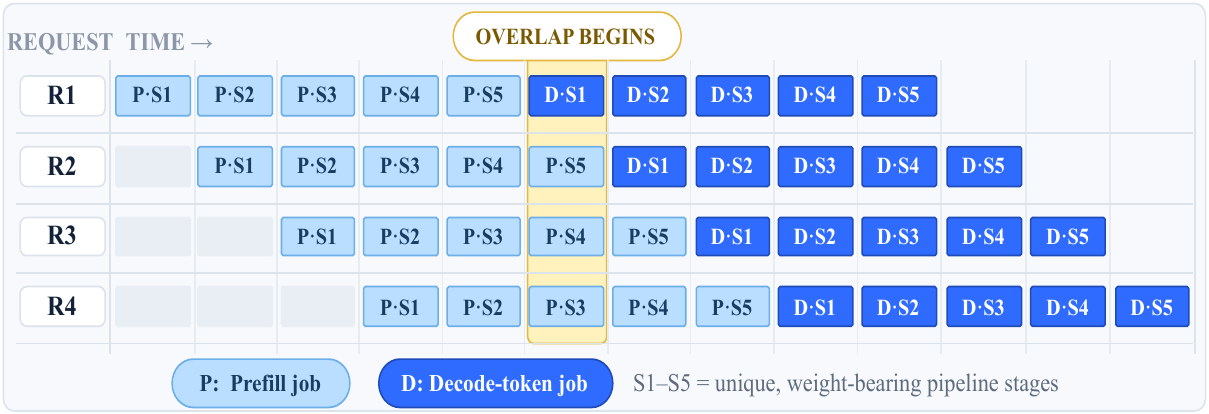}
    \caption{Scheduling of non-replicated multi-request pipeline.}
    \label{fig:non_replicated_pipeline}
\end{figure}

\subsection{Hierarchical KV Memory Allocation}

\textit{CHIPSMORE} employs a composable hierarchical \ac{KV}-memory allocation scheme that dynamically utilizes router scratchpad memory, \ac{SRAM-DCIM}, and \ac{eDRAM} according to the workload requirements and execution mode (Table~\ref{tab:op_mode}).
The memory hierarchy is organized according to both access locality and storage density.
Router scratchpad memory serves as the first-tier \ac{KV} storage due to its proximity to the \ac{IPCN} and \ac{PE}, thereby minimizing access and communication overheads during attention computation.
\ac{SRAM-DCIM} forms the second-tier storage resource, providing additional capacity while maintaining direct accessibility from the router.
The \ac{eDRAM} acts as a capacity-oriented third tier and is utilized only when the aggregated \ac{KV}-cache demand exceeds the combined capacity of the router scratchpad and \ac{SRAM-DCIM}.

The allocation policy is mode-dependent.
During base-mode inference, \ac{SRAM-DCIM} resources are not required for adaptation parameters and are therefore repurposed as \ac{KV} storage.
In contrast, during \ac{LoRA} inference, \ac{SRAM-DCIM} capacity is reserved for storing the low-rank adaptation matrices and is excluded from the \ac{KV}-storage pool.
This dynamic allocation policy enables efficient reuse of on-chip memory resources while supporting multiple operating modes without dedicated hardware provisioning.
The \ac{KV} data remain localized within the \ac{CT} whenever sufficient capacity is available, thereby reducing inter-chiplet communication overhead.
When the \ac{KV}-cache requirement exceeds the aggregate memory capacity of a single \ac{CT}, \textit{CHIPSMORE} distributes the \ac{KV} data across adjacent chiplets through capacity striping.

\subsection{Non-Replicated Multi-Request Pipeline}

Multi-request inference is commonly implemented by replicating model weights across multiple processing instances to increase throughput.
While such an approach exposes request-level parallelism, it scales the required weight-storage capacity approximately in proportion to the number of concurrent requests, leading to increased RRAM area, chiplet count, leakage power, and system cost.
To address this limitation, \textit{CHIPSMORE} adopts a non-replicated multi-request execution pipeline (\Cref{fig:non_replicated_pipeline}) that increases throughput without duplicating the pretrained weights stored in the \ac{RRAM-ACIM} arrays.

\begin{table}[t]
    \caption{CHIPSMORE System Parameter}
    \centering
    \setlength{\tabcolsep}{7pt}
    \begin{tabular}{c|c|c|c}
         \hline
         \multicolumn{4}{c}{\textbf{System Level}} \\
         \hline
         Bit-width & 64 & Frequency & 1 GHz \\
         \hline
         Tech. Node & 7 nm & Cluster size & 4 chiplets \\
         \hline \hline
         \multicolumn{4}{c}{\textbf{Compute Tile Level}} \\
         \hline
         IPCN Dimension & 32$\times$32 & Scratchpad Cap. & 16 MiB \\
         \hline
         SRAM-DCIM Cap. & 16 MiB & eDRAM Cap. & 64 MiB \\
         \hline
         UCIe endpoints & 2 & eDRAM Refresh & 10 ms \\
         \hline
         UCIe config. & \multicolumn{2}{c}{16 lanes / endpoint} & \\
         \hline \hline
         \multicolumn{4}{c}{\textbf{Macro Level (per unit Router-PE pair)}} \\
         \hline
         RRAM-ACIM \cite{rram_acim} & 256$\times$256 & SRAM-DCIM \cite{sram_dcim} & 256$\times$64 \\ 
         \hline
         Inter-Router I/O Ports & 4 & AXI-Stream (pairs) & 2 \\ 
         \hline
         FIFO Size (each) & 256 B & \\ 
         \hline
    \end{tabular}
    
    \vspace{0.15cm}
    {\raggedright *Cap.: capacity for KV caching\\ }
    
    \label{tab:sys_param}
\end{table}

The execution model leverages the layer-wise organization of the accelerator.
Each transformer layer is statically assigned to a unique weight-bearing \ac{CT} cluster.
During execution, multiple inference requests are injected into the pipeline at different temporal offsets.
As one request advances to the subsequent layer, the preceding layer immediately becomes available to process another request.
Consequently, several requests can occupy different layers of the model simultaneously, forming a pipeline across the distributed compute-tile clusters.
This scheduling strategy exploits the inherent sequential structure of transformer inference while maintaining a single physical copy of the pretrained parameters.

Under this pipeline, model weights are shared across all active requests, whereas the runtime states remain independent.
Specifically, each request maintains its own sequence context, KV cache, and generation state throughout execution.
Intermediate activations propagate through the layer pipeline, while request-specific KV data are retained within the hierarchical memory system described previously.
The throughput benefit is particularly significant during the decode phase.
Unlike the prefill phase, which already exhibits substantial intra-request parallelism, autoregressive decoding generates only a single token per request in each iteration.
Consequently, single-request execution underutilizes the distributed computational resources of the \ac{IPCN} and \acp{PE}.
By interleaving multiple requests across the layer pipeline, \textit{CHIPSMORE} converts request-level parallelism into hardware utilization, enabling otherwise idle compute tiles to process different requests concurrently.

\subsection{State-Aware Resource Reconfiguration}
The memory footprint of \textit{CHIPSMORE} contains multiple classes of state with distinct retention requirements.
Pretrained model parameters stored in \ac{RRAM-ACIM} are nonvolatile and remain intact without power.
In contrast, \ac{LoRA} parameters stored in \ac{SRAM-DCIM} and \ac{KV}-cache data residing in router scratchpad, \ac{SRAM-DCIM}, or \ac{eDRAM} constitute volatile runtime state whose contents must be preserved throughout task execution and request servicing.
Consequently, power management decisions cannot be based solely on compute activity, but must additionally account for state residency within the memory hierarchy.

To accommodate these heterogeneous retention requirements, \textit{CHIPSMORE} employs a state-aware resource reconfiguration policy that manages compute, communication, and memory subsystems.
During layer execution, the \ac{IPCN}, \ac{CIM} macros, and associated memory resources remain active to support inference computation.
Upon completion of execution, a \ac{CT} cluster transitions to a reduced-power state determined by the type of information retained locally.
For clusters containing \ac{LoRA} parameters or active \ac{KV}-cache entries, only the corresponding memory resources remain powered while \ac{IPCN} routers and inactive compute modules are power gated.
By decoupling memory retention from compute activation, \textit{CHIPSMORE} selectively enables only the hardware components required by the current workload phase, thereby reducing both dynamic and static power consumption while preserving the execution state required for subsequent requests.

Furthermore, the proposed mechanism allows system power to scale primarily with active computation and retained runtime state rather than with total model capacity.
Despite model size increases, only the clusters participating in the current execution stage require full activation.
As context length and request concurrency increase, additional power is incurred predominantly by the memory resources required for KV retention.

\section{System Evaluation}

\begin{figure}[t]
    \centering
    \includegraphics[width=0.9\linewidth]{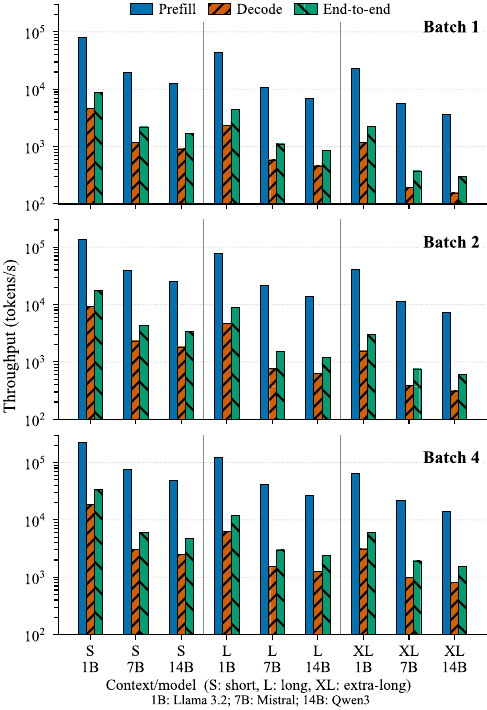}
    \caption{Throughput vs context length and batch size (base mode).}
    \label{fig:throughput_vs_batch_context}
\end{figure}

The proposed system was evaluated using a comprehensive hardware–software co‑verification methodology, with parameters in Table~\ref{tab:sys_param}.
Digital hardware components were implemented and functionally verified using Verilog HDL.
Logic synthesis was performed with \textit{Synopsys Design Compiler} with $7 nm$ technology node.
The power and area of SRAM scratchpad macros were obtained using \textit{CACTI}~\cite{cacti}.
The end‑to‑end \ac{LLM} inference was evaluated using a cycle‑accurate, instruction‑level simulator that implements the \ac{IPCN} instruction set along with the proposed mapping and dataflow schemes \cite{iccad_leap,iscas_primal}.
The simulator captures both computation and communication behavior across compute tiles, enabling detailed analysis of system‑level performance, energy efficiency, and scalability under representative \ac{LLM} workloads with INT8 weight configuration.

\subsection{Performance Benchmark}

\subsubsection{Throughput vs Model Size}
The throughput decreases with increasing model size and context length.
For a fixed workload, larger models require significantly more projection, attention, and feed-forward computations per generated token, while also increasing the depth of the layer pipeline.
Consequently, activations must traverse a larger number of weight-bearing \ac{CT} clusters before token generation can complete.
This effect is particularly evident for Qwen3-14B, which contains more transformer layers, larger hidden dimensions, and larger feed-forward networks than Mistral-7B, despite both models employing eight K/V heads with a head dimension of 128.
The increased computational workload lengthens the critical execution path through the \ac{IPCN} and \ac{PE} hierarchy, thereby reducing both prefill and decode throughput.

\subsubsection{Throughput vs Context Length}

\begin{figure}[t]
    \centering
    \includegraphics[width=\linewidth]{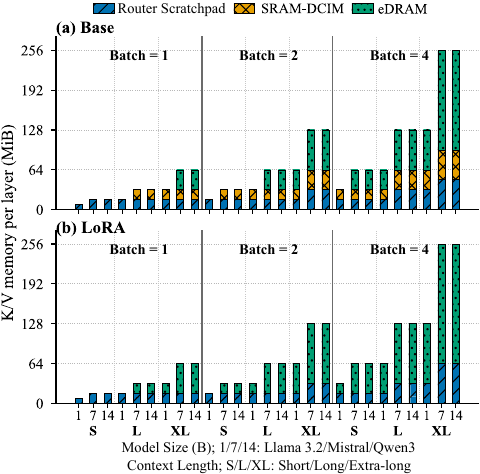}
    \caption{KV memory allocation on different modes, context lengths and batch sizes (per LLM layer).}
    \label{fig:kv_mem_allocation}
\end{figure}

Increasing the context length from S to XL also reduces throughput across all models.
This degradation originates from both computational and memory effects.
During the prefill phase, the attention computation scales quadratically with prompt length because the attention-score matrix grows as $O(T^2)$.
During autoregressive decoding, each newly generated token must attend to all previously retained tokens, causing the amount of K/V data accessed per decoding step to grow linearly with context length.
As a result, longer contexts increase both attention latency and memory traffic.
Furthermore, KV-cache capacity scales proportionally with retained sequence length.
For the XL workload, the \ac{KV} memory required exceeds the combined capacity of the router scratchpad and \ac{SRAM-DCIM}.
Consequently, the capacity-oriented \ac{eDRAM} tier is activated, introducing additional memory-access latency and interconnect traffic.
Although the hierarchical memory organization avoids off-chip accesses, the increased reliance on \ac{eDRAM} nevertheless lowers the achievable throughput.

\subsubsection{Throughput vs Batch Size}
Increasing batch size improves aggregate throughput by raising utilization of the non-replicated multi-request pipeline.
Multiple requests are temporally interleaved across the layer-wise \ac{CT} clusters, allowing different pipeline stages to process different requests concurrently while sharing a single physical copy of the pretrained weights in the \ac{RRAM-ACIM} arrays.
This mechanism converts request-level parallelism into hardware utilization without replicating model parameters. 
For the short workload, increasing batch size from 1 to 4 improves end-to-end throughput by 3.91× for Llama 3.2-1B, 2.71× for Mistral-7B, and 2.77× for Qwen3-14B.

The throughput improvement is nevertheless sub-linear with respect to batch size, particularly for the larger models.
Several factors contribute to this behavior.
First, pipeline fill and drain latency become increasingly significant as the number of transformer layers grows, limiting effective pipeline utilization. 
Second, individual pipeline stages maintain exclusive occupancy and therefore cannot process multiple requests simultaneously within the same layer.
Third, all requests share \ac{IPCN} communication resources, memory bandwidth, and \ac{KV}-cache access paths, leading to contention as concurrency increases.

\subsection{KV Memory Allocation}

\Cref{fig:kv_mem_allocation} illustrates the single-layer \ac{KV}-memory composition of \textit{CHIPSMORE} for both base-mode and \ac{LoRA} inference across different model sizes, context lengths, and batch sizes.
For both execution modes, the \ac{KV}-cache requirement increases proportionally with context length.
Consequently, long- and extra-long-context workloads increasingly stress the on-chip memory hierarchy despite the \ac{KV}-cache size remaining independent of the frozen model weights.

During base-mode inference, the \ac{SRAM-DCIM} macros are available as auxiliary \ac{KV} storage and therefore participate in the hierarchical allocation.
Short-context workloads fit entirely within the router scratchpad tier.
As the context length increases, \ac{KV} data progressively occupy the \ac{SRAM-DCIM} resources.
When the \ac{KV} requirement reaches 64 MiB per layer under the extra-long workload, the combined capacity of router scratchpad and \ac{SRAM-DCIM} is exceeded and the \ac{eDRAM} tier is activated.
The memory composition changes during \ac{LoRA} inference because \ac{SRAM-DCIM} capacity is reserved for the low-rank matrices.
As a result, \ac{KV} data are allocated only between router scratchpad and \ac{eDRAM}.
Compared with base-mode inference, \ac{LoRA} execution therefore increases dependence on \ac{eDRAM} at long contexts because \ac{SRAM-DCIM} can no longer contribute to \ac{KV} capacity.

Together, this demonstrates that the proposed hierarchical memory organization effectively absorbs the \ac{KV}-capacity growth caused by increasing context length while preserving locality within the \ac{CT}.
More importantly, the allocation policy dynamically repurposes \ac{SRAM-DCIM} resources when \ac{LoRA} support is not required, thereby maximizing effective on-chip memory capacity and reducing the need for external memory resources.

\subsection{Performance Comparison of LoRA vs Base Modes}

\begin{table}[t]
    \caption{End-to-End Throughput: LoRA Mode}
    \centering
    \begin{tabular}{c|c|c|c|c|c}
         \hline
         \multicolumn{2}{c|}{} & \multicolumn{2}{c|}{Batch Size 1} & \multicolumn{2}{c}{Batch Size 4}\\
         \hline
         Context & Model$^a$ & Throughput & Perf. & Throughput & Perf. \\
         Length & & (tokens/s) & Factor$^b$ & (tokens/s) & Factor$^b$ \\  
         \hline
         \multirow{3}{*}{S}
         & 1B & 8767.65 & $0.99 \times$ & 23501.00 & $0.71 \times$ \\
         & 7B & 2161.53 & $0.98 \times$ & 5038.49 & $0.85 \times$ \\
         & 14B & 1619.09 & $0.95 \times$ & 3996.63 & $0.84 \times$ \\
         \hline
         \multirow{3}{*}{L}
         & 1B & 4401.97 & $0.99 \times$ & 10034.14 & $0.86 \times$ \\
         & 7B & 752.46 & $0.71 \times$ & 2853.67 & $0.95 \times$ \\
         & 14B & 596.40 & $0.72 \times$ & 2287.2 & $0.96 \times$ \\
         \hline
         \multirow{3}{*}{XL}
         & 1B & 1510.22 & $0.72 \times$ & 5602.30 & $0.94 \times$ \\
         & 7B & 318.53 & $0.84 \times$ & 1850.69 & $0.96 \times$ \\
         & 14B & 253.00 & $0.85 \times$ & 1466.35 & $0.96 \times$ \\
         \hline
    \end{tabular}
    
    \vspace{0.15cm}    
    {\raggedright \scriptsize $^a$ $1B/7B/14B: Llama 3.2/Mistral/Qwen3$ \\}
    \vspace{0.05cm}
    {\raggedright \scriptsize $^b$ $Perf.$ $Factor = Throughput_{LoRA}/Throughput_{Base}$ \\}
    \label{tab:benchmark_tp_pwr}
\end{table}

Table~\ref{tab:benchmark_tp_pwr} reports the end-to-end throughput of \textit{CHIPSMORE} in \ac{LoRA} mode using rank-8 adaptation on the Q and V projections.
Overall, \ac{LoRA} throughput remains close to that of the base model across most operating points, indicating that the overhead introduced by the low-rank adaptation computation itself is relatively small.
Instead, the dominant source of performance degradation originates from changes in the \ac{KV}-memory allocation policy, as shown in \Cref{fig:kv_mem_allocation}.

For long-context workloads, this effect is particularly evident in the larger models.
Under base-mode inference, Mistral-7B and Qwen3-14B can be fully accommodated using router scratchpad and \ac{SRAM-DCIM} without eDRAM accesses.
In contrast, \ac{LoRA} execution reserves the \ac{SRAM-DCIM} resources for adaptation weights, forcing half of the \ac{KV} data into \ac{eDRAM}.
This additional \ac{eDRAM} traffic increases the latency of \ac{KV} retrieval during both prefill and decode attention operations, reducing throughput to $\sim$71\% of the corresponding base-mode performance for batch size one.
The impact is substantially smaller at batch size four, where the performance factor increases to 0.95$\times$ because the larger amount of request-level parallelism improves pipeline utilization and amortizes the additional memory latency across multiple active requests.

Llama 3.2-1B exhibits a distinct behavior due to its smaller \ac{KV}-cache footprint.
For batch size one, both the short- and long-context workloads fit entirely within router scratchpad in both execution modes, resulting in negligible performance differences. 
The impact of hierarchical-memory allocation only appears at the extra-long context. 
In this case, base-mode inference utilizes both router scratchpad and \ac{SRAM-DCIM}, whereas \ac{LoRA} must allocate the overflow capacity in \ac{eDRAM}.
The resulting increase in \ac{KV}-access latency reduces the throughput ratio to 0.72$\times$.
However, as the batch size increases to four, the multi-request pipeline improves the performance factor to 0.94$\times$.

These results demonstrate that the throughput overhead of \ac{LoRA} execution in \textit{CHIPSMORE} is primarily determined by resource contention within the \ac{KV}-memory system rather than by the computational cost of the low-rank adaptation itself.
The proposed memory hierarchy therefore enables efficient support for \ac{LoRA} inference, with the throughput penalty remaining $\sim$5\% for most long-context and high-concurrency operating points.

\subsection{Pipeline Utilization}

\begin{figure}[t]
    \centering
    \includegraphics[width=\linewidth]{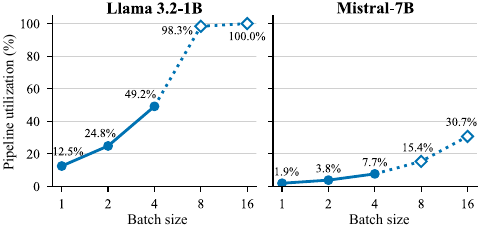}
    \caption{Pipeline utilization across various batch size.}
    \label{fig:pipeline_utilization}
\end{figure}

The average utilization of the layer-wise execution pipeline for Llama 3.2-1B and Mistral-7B as the batch size increases is shown in \Cref{fig:pipeline_utilization}.
The utilization is defined as the percentage of weight-bearing \ac{CT} clusters actively executing inference requests over the total number of \ac{CT} clusters allocated to the model.
For both models, pipeline utilization increases with batch size because the non-replicated multi-request execution scheme injects independent requests into different stages of the layer pipeline.
As the number of concurrent requests increases, idle pipeline stages are progressively filled, improving utilization of the distributed \acp{CIM} and \ac{IPCN} resources without replicating pretrained weights.

Llama 3.2-1B exhibits rapid utilization growth.
This behavior is a direct consequence of the model's relatively shallow transformer stack consisting of 16 layers.
Since each layer is mapped to a dedicated weight-bearing pipeline stage, a batch size comparable to the pipeline depth is sufficient to occupy almost all stages.
Once the pipeline becomes fully populated, further increases in batch size provide little additional utilization benefit because all stages are already active.
In contrast, Mistral-7B achieves lower utilization.
This originates from the significantly deeper execution pipeline resulting from its 32-layer transformer architecture and the increased number of \ac{CT} clusters required to accommodate the larger model.
Although additional requests improve occupancy, the available degree of request-level parallelism remains insufficient to fully populate the deeper pipeline.
The results highlight an important characteristic of the proposed non-replicated execution model.
Throughput improvements are obtained by increasing pipeline occupancy rather than by duplicating model weights.
The achievable utilization is bounded by the relationship between batch size and pipeline depth.

\begin{figure}[t]
    \centering
    \includegraphics[width=0.9\linewidth]{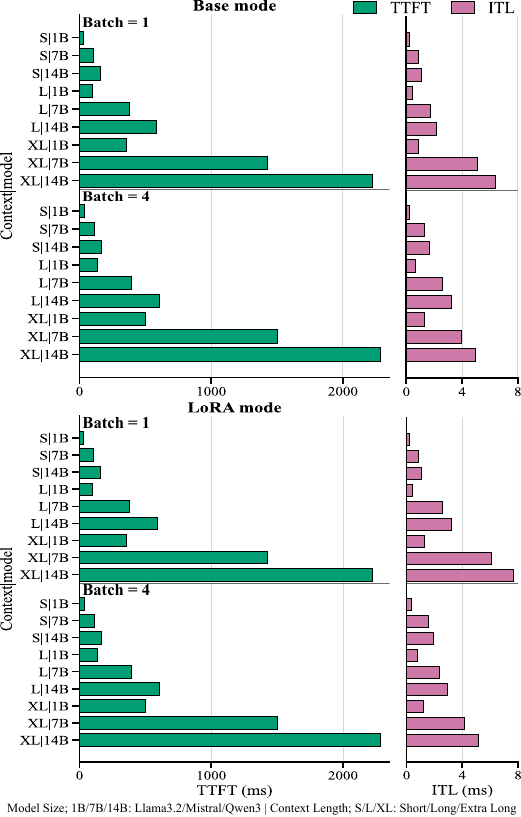}
    \caption{TTFT and ITL on different modes, context lengths and batch sizes.}
    \label{fig:ttft_itl}
\end{figure}

The existence of underutilized stages in deep-model workloads does not translate directly into proportional power consumption.
Under the proposed state-aware resource reconfiguration scheme, \ac{CT} clusters that are not actively participating in computation can transition into reduced-power states while retaining only the memory resources required for \ac{LoRA} parameters or \ac{KV}-cache data.
Consequently, the idle pipeline capacity observed in \Cref{fig:pipeline_utilization} can be effectively exploited for power reduction through state-aware power gating, which is analyzed in the following subsection.

\subsection{\ac{TTFT} and \ac{ITL}}

\Cref{fig:ttft_itl} shows the batch-averaged \ac{TTFT} and effective \ac{ITL}.
\ac{TTFT} is dominated by prompt processing and increases substantially with model size and context length.
Increasing the prompt length from short to extra-long raises TTFT by 14$\times$, reflecting the quadratic attention workload and increased communication.
Increasing the batch size from 1 to 4 raises \ac{TTFT} by $\sim$20\%, substantially below linear batch scaling due to interleaved pipeline execution. 
\ac{LoRA} increases \ac{TTFT} $< 5\%$, indicating that its programming and adapter operations are largely hidden behind the prefill critical path.
\ac{ITL} generally increases with model size and context length as each autoregressive step requires more \ac{KV}-cache access, attention processing, and inter-tile communication.
Its batch dependence is non-proportional, where additional requests improve pipeline occupancy and amortize fill/drain overhead.
\ac{LoRA} increases \ac{ITL} when adapter computation enters the decode critical path. 
However, the penalty can be partially hidden by batched operations and pipeline scheduling, particularly at batch size four.

\subsection{Inter-CT UCIe Traffic and Utilization}

\begin{figure}[t]
    \centering
    \includegraphics[width=0.85\linewidth]{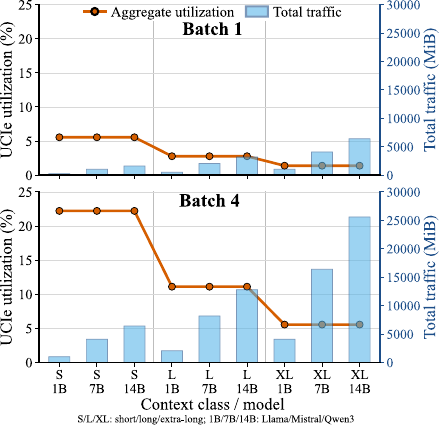}
    \caption{Inter-CT UCIe traffic and utilization on different models, context lengths and batch sizes.}
    \label{fig:ucie_util}
\end{figure}

\Cref{fig:ucie_util} summarizes the aggregate inter-chiplet traffic and corresponding \ac{UCIe} utilization for different model sizes, context lengths, and batch sizes.
The total traffic represents all inter-\ac{CT} data transfers carried by the \ac{UCIe} channel, while utilization is measured relative to the available \ac{UCIe} bandwidth over the entire execution period.

\begin{figure}[t]
    \centering
    \includegraphics[width=\linewidth]{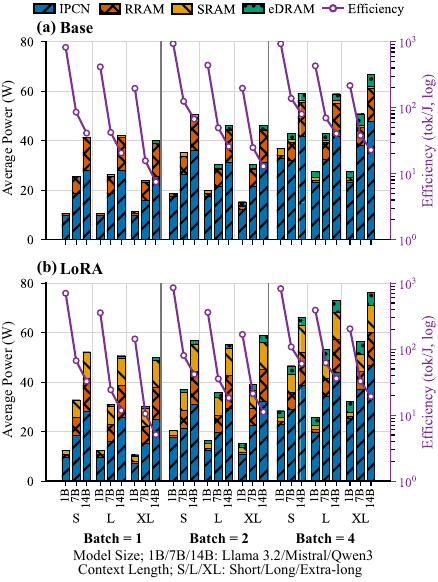}
    \caption{Average system power and energy efficiency on different modes, context lengths and batch sizes.}
    \label{fig:power_efficiency}
\end{figure}

Total traffic increases with model size and batch size because larger models require more chiplets and deeper layer pipelines, while additional requests generate more concurrent activation and \ac{KV}-related transfers.
Consequently, Qwen3-14B exhibits the highest traffic volume among the evaluated models.
In contrast, \ac{UCIe} utilization decreases with increasing context length.
Although longer contexts generate more inter-chiplet traffic, the execution time increases more rapidly due to the quadratic growth of prefill attention, larger \ac{KV}-cache accesses during decode, and increased intra-chiplet \ac{IPCN} activity.
As a result, \ac{UCIe} transfers are amortized over a longer execution interval, lowering average link occupancy.
This explains why extra-long workloads exhibit lower utilization despite generating higher traffic volumes.

Even at batch size four, the maximum observed \ac{UCIe} utilization remains below 25\%, indicating that the inter-chiplet fabric is not a system bottleneck.
These results confirm that \textit{CHIPSMORE} localizes the majority of data movement within each compute tile, while the \ac{UCIe} interconnect provides sufficient bandwidth to support larger models, longer contexts, and higher request concurrency without saturation.

\subsection{System Power Scaling and Energy Efficiency}

\begin{table*}[t]
    \caption{Comparison with SoTA Platforms}
    \centering
    \setlength{\tabcolsep}{6.5pt}
    \begin{tabular}{c|c|c|c|c|c|c|c|c}
         \toprule
         \multicolumn{2}{c|}{Platform} & Nvidia H100 & Apple M4-Max & Cerebras-2 \cite{cerebras_wse} & CENT \cite{cent} & H2-LLM \cite{h2_llm} & CHIME \cite{chime} & \textbf{This Work} \\
         \hline
         \multirow{5}{*}{\shortstack{Batch\\Size 1}} 
         & Throughput (tokens/s) & 467.3 & 90.7 & 4271.5 & 1230.3 & 235.4 & 1392.0 & 1112.5 \\
         & Avg. Power (W) & 350 & 55.0 & 18000 & 220 & 23.5 & 2215 & 30.7 \\
         & Efficiency (tokens/J) & 1.34 & 1.65 & 0.24 & 5.59 & 10.01 & 0.63 & 36.24 \\
         & Speedup & $1\times$ & $0.19\times$ & $9.14\times$ & $2.63\times$ & $0.5\times$ & $2.98\times$ & $2.38\times$ \\
         & Efficiency$\times$ & $1\times$ & $1.24\times$ & $0.18\times$ & $4.19\times$ & $7.49\times$ & $0.47\times$ & $27\times$\\

         \hline
         \multirow{5}{*}{\shortstack{Batch\\Size 4}} 
         & Throughput (tokens/s) & 1665.7 & 280.4 & 8771.6 & 3124.9 & 822.2 & 6516.2 & 3003.9 \\
         & Avg. Power (W) & 490 & 70 & 20700 & 240 & 29.3 & 2600 & 46.4 \\
         & Efficiency (tokens/J) & 3.40 & 4.01 & 0.42 & 13.02 & 28.02 & 2.51 & 64.73 \\
         & Speedup & $1\times$ & $0.17\times$ & $5.27\times$ & $1.88\times$ & $0.49\times$ & $3.91\times$ & $1.80\times$\\
         & Efficiency$\times$ & $1\times$ & $1.18\times$ & $0.12\times$ & $3.83\times$ & $8.24\times$ & $0.74\times$ & $19.04\times$ \\
         
         \bottomrule
    \end{tabular}
    
    \vspace{0.1cm}
    {\raggedright *Evaluations based on Mistral-7B-INT8, base-mode long context (4096/4096) $\vert$ Baseline: Nvidia H100 with vLLM \\ }
    
    \label{tab:cross_platform}
\end{table*}

The system average power and end-to-end energy efficiency are reported in \Cref{fig:power_efficiency}.
Across all evaluated configurations, the \ac{IPCN} and router subsystem remains the dominant contributor to overall power consumption.
This behavior is expected because the \ac{IPCN} simultaneously performs inter-router communication, collective operations, and \ac{DMAC} computations during inference execution.

\begin{figure}[t]
    \centering
    \includegraphics[width=0.7\linewidth]{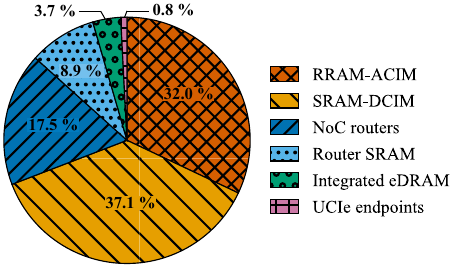}
    \caption{Area breakdown per compute tile.}
    \label{fig:area_breakdown}
\end{figure}

Average power generally increases with both model size and batch size.
Larger models require deeper computation pipelines, resulting in more concurrently active \ac{IPCN} routers and \ac{CIM} resources.
Similarly, increasing batch size improves pipeline occupancy, thereby activating a larger fraction of the distributed compute resources.
The relationship between power and context length is more nuanced.
Although longer contexts always increase the total amount of computation, communication, and \ac{KV}-cache accesses, average power does not scale proportionally with context length because power is calculated as total energy divided by execution time.
When the \ac{KV} footprint spills into \ac{eDRAM}, execution latency often increases faster than computation intensity.
As a result, a long-context workload may consume substantially more total energy while exhibiting only a modest increase, or even a slight reduction, in average power.

\ac{LoRA} mode consistently consumes higher power than that of the corresponding base-mode configuration.
In base-mode inference, \ac{SRAM-DCIM} resources are repurposed as auxiliary \ac{KV} storage and are accessed only during \ac{KV} retrieval.
During LoRA inference, however, \ac{SRAM-DCIM} resources remain allocated to adaptation weights while participating in \ac{LoRA} computations. 
Consequently, both \ac{SRAM-DCIM} activity and \ac{eDRAM} utilization increase.

The energy-efficiency trends closely follow the interaction between throughput and power.
Increasing batch size generally improves energy efficiency because the non-replicated multi-request pipeline amortizes pipeline fill and drain overhead over a larger number of requests while improving utilization of the weight-bearing \ac{CT} clusters.
Consequently, throughput increases faster than power consumption across most operating points.
In contrast, efficiency decreases with model size and context length due to joint effects of lower throughput and increasing power consumption, caused by increasing weight-pipeline depth, communication traffic, and \ac{KV} service time.

Although larger models and longer contexts inevitably increase the computational and memory demands of inference, the power scaling remains significantly more favorable than throughput scaling, enabling efficient operation across a broad range of deployment scenarios.

\subsection{Comparison with SoTA}

Table~\ref{tab:cross_platform} compares \textit{CHIPSMORE} against representative GPU-, SoC-, wafer-scale-, and \ac{CIM}-based LLM inference platforms. 
\textit{CHIPSMORE} achieves 1112.5 tokens/s and 3003.9 tokens/s for batch sizes 1 and 4, respectively, while consuming only 30.7 W and 46.4 W of average power.
Although its absolute throughput is lower than highly parallel wafer-scale system such as Cerebras-2, \textit{CHIPSMORE} delivers substantially higher energy efficiency due to the combined benefits of \ac{CIM} execution, in-network computation, hierarchical KV-memory management, and non-replicated multi-request scheduling.
Beyond throughput and efficiency, \textit{CHIPSMORE} supports workload features that are not jointly available in prior \ac{CIM}-based \ac{LLM} accelerators, including \ac{LoRA} inference, hierarchical \ac{KV}-memory management, dynamic \ac{KV}-capacity scaling, and weight-replication-free multi-request execution.
These mechanisms enable operation across different adaptation modes, context lengths, and request concurrency levels within a unified architecture. 
Overall, the results suggest that \textit{CHIPSMORE} provides a competitive trade-off between throughput, workload flexibility, and energy efficiency for \ac{LLM} inference.

\section{Conclusion}
This paper presented \textit{CHIPSMORE}, a multi-mode and multi-request \ac{CIM} accelerator for efficient \ac{LLM} inference across diverse operating scenarios, including base-mode execution, LoRA adaptation, varying context lengths, and concurrent multi-request serving.
Building upon the compute-in-interconnect and \ac{CIM} paradigm, \textit{CHIPSMORE} integrates heterogeneous \ac{RRAM-ACIM} and \ac{SRAM-DCIM} \acp{PE} with a programmable \ac{IPCN} to support persistence-aware compute partitioning and in-network computation.
A hierarchical KV-memory organization dynamically orchestrates router scratchpad, \ac{SRAM-DCIM}, and \ac{eDRAM} resources to accommodate workload-dependent \ac{KV}-cache requirements, while a non-replicated multi-request pipeline improves hardware utilization without duplicating pretrained weights.
Furthermore, a state-aware resource reconfiguration mechanism reduces power consumption by selectively retaining runtime states.
Evaluation results demonstrate that \textit{CHIPSMORE} efficiently supports both base-mode and \ac{LoRA} inference while providing scalable \ac{KV}-cache management, replication-free multi-request execution, and favorable power scaling.
Compared with Nvidia H100, \textit{CHIPSMORE} achieves up to $2.38\times$ higher throughput and $27\times$ higher energy efficiency on Mistral-7B inference while eliminating weight replication for multi-request serving.
By jointly addressing model adaptation, \ac{KV}-cache scalability, request concurrency, and power-aware resource management, \textit{CHIPSMORE} establishes a unified and scalable architecture for future LLM inference systems that operates under stringent performance, capacity, and energy constraints.

\bibliographystyle{IEEEtran}
\bibliography{references}

\end{document}